\documentclass[12pt]{article}
 \usepackage{epsfig}
 \def\be{\begin{equation}}
 \def\ee{\end{equation}}
 \def\bea{\begin{eqnarray}}
 \def\eea{\end{eqnarray}}
 \usepackage{graphicx}

 \catcode`\@=11
 \def\lsim{\mathrel{\mathpalette\@versim<}}
 \def\gsim{\mathrel{\mathpalette\@versim>}}
 \def\@versim#1#2{\vcenter{\offinterlineskip
 \ialign{$\m@th#1\hfil##\hfil$\crcr#2\crcr\sim\crcr } }}
 \catcode`\@=12

 \catcode`@=12
\begin{document}
 \thispagestyle{empty}
 \begin{flushright}
 UCRHEP-T607\\
 Jan 2021\
 \end{flushright}
 \vspace{0.6in}
 \begin{center}
 {\LARGE \bf Hierarchical Quark and Lepton Masses\\ 
 with Sequential U(1) Gauge Symmetry\\}
 \vspace{1.2in}
 {\bf Ernest Ma\\}
 \vspace{0.2in}
{\sl Physics and Astronomy Department,\\ 
University of California, Riverside, California 92521, USA\\}
\end{center}
 \vspace{1.2in}

\begin{abstract}\
Instead of right-handed neutrino singlets, the standard model is extended 
to include lepton triplets $(\Sigma^+,\Sigma^0,\Sigma^-)$.  Each quark and 
lepton family may now transform under an anomaly-free $U(1)_X$ gauge symmetry, 
known already for many years.  A new sequential application is presented, 
using just the one Higgs doublet of the standard model, together with 
two $U(1)_X$ Higgs singlets.  The resulting structure has hierarchical 
quark and lepton masses, as well as a viable seesaw neutrino mass matrix. 
\end{abstract}

 \newpage
 \baselineskip 24pt
\noindent \underline{\it Introduction}~:~
The standard $SU(3)_C \times SU(2)_L \times U(1)_Y$ gauge model (SM) of 
quarks and leptons is anomaly-free~\cite{bim72}.  So is its conventional 
left-right gauge extension to 
$SU(3)_C \times SU(2)_L \times SU(2)_R \times U(1)_X$, with the $U(1)_X$ 
charge identified~\cite{d79,mm80} as $(B-L)/2$, where $B=1/3$ is baryon 
number and $L=1$ is lepton number.  Without $SU(2)_R$, an extra $U(1)_X$ 
gauge symmetry may be added to the SM, where it could be made anomaly-free 
under the condition~\cite{kmpz17}
\begin{equation}
\sum^3_{i=1} 3n_i + n'_i = 0,
\end{equation}
where $n_i,n'_i$ are the $U(1)_X$ quark and lepton charges of each family, 
assuming also the existence of the right-handed singlet neutrino $\nu_R$.
If $n_i=1/3$ and $n'_i=-1$, the well-known $B-L$ gauge symmetry is obtained. 
If family universality is abandoned, there are obviously many more solutions. 
For example, if $n_i=1/3$ and $n'_{1,2,3} = (0,0,-3)$, then $B-3L_\tau$ is 
realized~\cite{m98,ms98,mr98,bhmz10}.  Some other choices are discussed in 
Ref.~\cite{kmpz17}.

There is another possibility for $U(1)_X$ to be anomaly-free, which is much 
less known.  Instead of $\nu_R$, the lepton triplet 
$(\Sigma^+,\Sigma^0,\Sigma^-)_R$ is now added per family.  Neutrinos then 
acquire Majorana masses from $\Sigma^0_R$~\cite{flhj89} through the so-called  
Type III~\cite{m98-1} seesaw mechanism.  Remarkably, this rather awkward 
addition to the SM also accommodates an unusual $U(1)_X$ which is 
anomaly-free.  In this paper, a sequential application of this gauge 
symmetry to the three families of quarks and leptons is made, resulting 
in hierarchical quark and lepton masses, as well as realistic quark and 
neutrino mixing matrices.

\noindent \underline{\it Unusual $U(1)_X$}~:~
Consider $SU(3)_C \times SU(2)_L \times U(1)_Y \times U(1)_X$ as a possible 
extension of the SM, under which each family of quarks and leptons transform 
as follows:
\begin{eqnarray}
&& (u,d)_L \sim (3,2,1/6;n_1), ~~~ u_R \sim (3,1,2/3;n_2), ~~~ d_R \sim 
(3,1,-1/3;n_3), \nonumber \\ 
&& (\nu,e)_L \sim (1,2,-1/2;n_4), ~~~ e_R \sim (1,1,-1;n_5), ~~~ 
\Sigma_R \sim (1,3,0;n_6).
\end{eqnarray}
It has been shown~\cite{m02,bbb86} that $U(1)_X$ is free of all anomalies 
for the following assignments:
\begin{equation}
n_2 = {1 \over 4} (7n_1-3n_4), ~~~ n_3 = {1 \over 4} (n_1+3n_4), ~~~ 
n_5 = {1 \over 4} (-9n_1+5n_4), ~~~ n_6 = {1 \over 4} (3n_1+n_4),
\end{equation}
where $n_{1,4}$ are arbitrary except that $n_4 \neq -3n_1$ is required.  
This unusual gauge symmetry has been studied 
somewhat~\cite{mr02,bd05,aem09,abm16}, but always in the framework that 
$n_{1,2,3,4,5}$ are the same for each family.  In this paper, they are 
allowed to be different, choosing $n_1=0$ for all three, but $n_4=0,1,2$ 
for the third, second, first families respectively.

\noindent \underline{\it Higgs Sector}~:~
Let $\Phi = (\phi^+,\phi^0)$, then it should have $U(1)_X$ charges 
\begin{equation}
n_2-n_1 = n_1-n_3 = n_6-n_4 = {3 \over 4}(n_1-n_4), ~~~ 
n_4-n_5 = {1 \over 4} (9n_1-n_4).
\end{equation}
For $n_4 \neq -3n_1$, there must be two distinct Higgs doublets.  
For example, if $n_1=n_4$, then $\Phi_1$ has zero $U(1)_X$ charge, 
and $\Phi_2$ has charge $2n_1$.
The lepton triplet $\Sigma_R$ must also acquire a mass, hence a Higgs 
singlet $\eta$ with $U(1)_X$ charge $(3n_1+n_4)/2$ is needed, which 
may also serve to break the $U(1)_X$ gauge symmetry.

Once family universality is not imposed, there are other solutions.  There 
is no simple formula as Eq.~(1).  In this paper, $n_1$ is assigned zero 
for all three families, whereas $n_4$ takes on the values $(2,1,0)$ for 
the first, second, third families respectively.  The Higgs sector is 
chosen to have just one doublet with $U(1)_X$ charge zero as in the SM, 
and two singlets $\eta_{1,2}$ with charges $1/4$, $3/4$ respectively.  
Together they will result in hierachical quark and lepton 
masses with realistic quark and neutrino mixing matrices as shown below. 
The Higgs potential is easily contructed, containing in particular the 
term $\eta_2^* \eta_1^3$ which ensures that there is no broken 
global $U(1)$ symmetry.

\noindent \underline{\it Quark and Lepton Sector}~:~
Under $U(1)_X$, the quarks and leptons transform as:
\begin{eqnarray}
&& (u,d)_L, (c,s)_L, (t,b)_L \sim 0, ~~~ u_R,c_R,t_R \sim -3/2,-3/4,0, \\ 
&& d_R,s_R,b_R \sim 3/2,3/4,0, ~~~ (\nu_e,e)_L, (\nu_\mu,\mu)_L, 
(\nu_\tau, \tau)_L \sim 2,1,0, \\ 
&& e_R,\mu_R,\tau_R \sim 5/2,5/4,0, ~~~ 
\Sigma^e_R,\Sigma^\mu_R,\Sigma^\tau_R \sim 1/2,1/4,0.
\end{eqnarray}
Hence the third family of quarks and leptons obtain tree-level masses 
from coupling to the one Higgs doublet $\Phi$ as in the SM.  For the second 
family, 
\begin{equation}
\bar{c}_L c_R \sim \phi^0 \eta_2^*, ~~~ \bar{s}_L s_R \sim \bar{\phi}^0 
\eta_2, ~~~ \bar{\mu}_L \mu_R \sim \bar{\phi}^0 \eta_1, ~~~ 
\bar{\nu}_\mu \Sigma^\mu \sim \phi^0 \eta_2^*.
\end{equation}
Their masses thus come from dimension-five operators, proportional to 
$v_0 v_{1,2}/\Lambda$, where $v_0 = \langle \phi^0 \rangle$, 
$v_1 = \langle \eta_1 \rangle$, $v_2 = \langle \eta_2 \rangle$.
Similarly, for the first family,
\begin{eqnarray}
\bar{u}_L u_R \sim \phi^0 \eta_2^* \eta_2^*, ~~~ 
\bar{d}_L d_R \sim \bar{\phi}^0 \eta_2 \eta_2,  
~~~ \bar{e}_L e_R \sim \bar{\phi}^0 \eta_1 \eta_1, \bar{\phi}^0 
\eta_2 \eta_1^*, ~~~  
\bar{\nu}_e \Sigma^e \sim \phi^0 \eta_2^* \eta_2^*.
\end{eqnarray}
Dimension-six operators are now required for their masses.  This pattern 
is maintained by the $U(1)_X$ charge assignments of the singlets $\eta_{1,2}$. 

The $3 \times 3$ quark mass matrices linking the left-handed doublets to 
the right-handed singlets are then of the form
\begin{equation}
{\cal M}_u = \pmatrix{ m_{uu} & 0 & 0 \cr m_{cu} & m_{cc} & 0 \cr m_{tu} & 
m_{tc} & m_{tt}}, ~~~ 
{\cal M}_d = \pmatrix{ m_{dd} & 0 & 0 \cr m_{ds} & m_{ss} & 0 \cr m_{bd} & 
m_{bs} & m_{bb}}. 
\end{equation}
This pattern is the result of the fact that all left-handed quarks have 
the same $n_1=0$.  After diagonalizing on the left by $U^\dagger_L(u,d)$ 
and on the right by $U_R(u,d)$, the quark mixing matrix is then given by 
$U_{CKM} = U_L^\dagger(u) U_L(d)$.

In the lepton sector, taking into account the off-diagonal terms
\begin{equation}
\bar{e}_L \mu_R \sim \bar{\phi}^0 \eta_2^*, ~~~ 
\bar{\mu}_L e_R \sim \bar{\phi}^0 \eta_2 \eta_2,  
\end{equation}
and ignoring higher-dimensional contributions, the $3 \times 3$ 
charged-lepton mass matrix is of the form
\begin{equation}
{\cal M}_l = \pmatrix{ m_{ee} & m_{e \mu} & 0 \cr m_{\mu e} & m_{\mu \mu} & 
0 \cr 0 & 0 & m_\tau}.
\end{equation}
Similarly, taking into account the off-diagonal terms
\begin{equation}
\bar{\nu}_\mu \Sigma^e \sim \phi^0 \eta_1^* \eta_1^*, \phi^0 \eta_2^* \eta_1, 
~~~ 
\bar{\nu}_\mu \Sigma^\tau \sim \phi^0 \eta_1^* \eta_2^*, ~~~ 
\bar{\nu}_\tau \Sigma^e \sim \phi^0 \eta_1 \eta_1, \phi^0 \eta_2 \eta_1^*, ~~~ 
\bar{\nu}_\tau \Sigma^\mu \sim \phi^0 \eta_1,
\end{equation}
and ignoring higher-dimensional contributions, the $3 \times 3$ 
neutrino mass matrix linking $\nu_L$ to $\Sigma_R$ is of the form
\begin{equation}
{\cal M}_D = \pmatrix{ m'_{ee} & 0 & 0 \cr m'_{\mu e} & m'_{\mu \mu} & 
m'_{\mu \tau} \cr m'_{\tau e} & m'_{\tau \mu} & m'_{\tau \tau}}.
\end{equation}
Finally the $3 \times 3$ Majorana mass matrix of $\Sigma$ is given by
\begin{equation}
{\cal M}_\Sigma = \pmatrix{M_{ee} & M_{e \mu} & M_{e \tau} \cr M_{\mu e} & 
M_{\mu \mu} & M_{\mu \tau} \cr M_{\tau e} & M_{\tau \mu} & M_{\tau \tau}},
\end{equation}
where $M_{\tau \tau}$ is an allowed invariant mass, and
\begin{eqnarray}
&& M_{\tau \mu} = M_{\mu \tau} \sim \eta_1, ~~~ 
M_{\mu e} = M_{e \mu} \sim \eta_2, ~~~ 
M_{\mu \mu} \sim \eta_1 \eta_1, \eta_2 \eta_1^*, \nonumber \\  
&& M_{\tau e} = M_{e \tau} \sim \eta_1 \eta_1, \eta_2 \eta_1^*, ~~~ 
M_{e e} \sim \eta_1 \eta_2.
\end{eqnarray}
Together, the three matrices allow for realistic charged lepton and neutrino 
masses with the correct $3 \times 3$ mixing matrix.  To see this, let 
both ${\cal M}_l$ and ${\cal M}_D$ be diagonal, with 
$M_{\tau e} = M_{\mu \mu} = M_{e \tau} = 0$ in ${\cal M}_\Sigma$, then 
there are two cofactor zeros~\cite{l05,m05} in the seesaw neutrino mass 
matrix, corresponding to the Y1 pattern studied in Ref.~\cite{lmw14}. 
This is known to have an acceptable fit to the neutrino data.  Relaxing 
this assumption will only improve the fit.

\noindent \underline{\it Scotogenic Muon Mass}~:~
As an example, the muon mass may be generated by a dimension-five operator 
in one loop through dark matter.  This scotogenic mechanism is shown in 
Fig.~1, where
\begin{figure}[htb]
\vspace*{-5cm}
\hspace*{-3cm}
\includegraphics[scale=1.0]{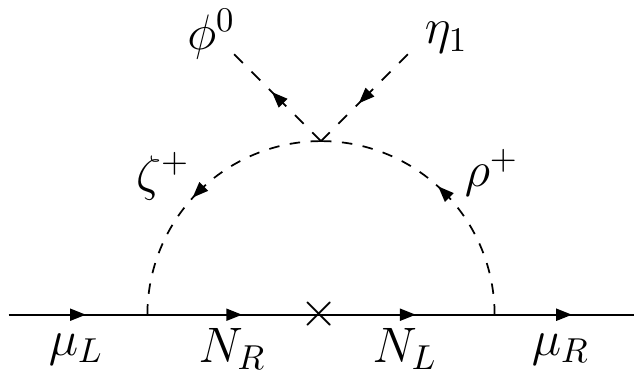}
\vspace*{-21.5cm}
\caption{Scotogenic $\mu$ mass.}
\end{figure}
\begin{equation}
N_{L,R} \sim -1/4, ~~~ (\zeta^+,\zeta^0) \sim -5/4, ~~~ \rho^+ \sim -3/2.
\end{equation}
Because of their $U(1)_X$ assignments, these new particles possess an 
automatic global $U(1)_D$, under which they transform together nontrivially, 
whereas all other fields are trivial.  The neutral Dirac fermion $N$ is then 
a dark-matter candidate.

This example is similar to recent proposals~\cite{m20,m21} of scotogenic 
lepton or quark mass, the difference being the $U(1)_X$ symmetry as the 
instigator of this mechanism.  Again, because of the radiative $\mu$ mass, 
the Higgs coupling to $\bar{\mu} \mu$ is predicted to be 
different~\cite{fm14,fmz16} from the SM result of $m_\mu/(246~{\rm GeV})$. 
Recently, ATLAS reports~\cite{atlas20} an observation of the $\mu^+ \mu^-$ 
mode at the level $1.2 \pm 0.6$ relative to the SM prediction.  Also, CMS 
has the result~\cite{cms20} $1.2 \pm 0.4 \pm 0.2$.  

The same diagram which generates the radiative $\mu$ mass also yields 
an anomalous magnetic moment, which is not suppressed by the usual $16 \pi^2$ 
loop factor, because it is now common to both evaluations~\cite{fmz16}.  
Furthermore, it is of the correct (positive) sign for the known discrepancy 
of the experimental measurement versus the SM theoretical prediction, 
despite the fact that the internal charged particles are scalar, as 
explained in Ref.~\cite{fmz16}.

\noindent \underline{\it Gauge Interactions}~:~
Since $U(1)_X$ assignments are different for each family, the couplings of 
the gauge boson $X$ are not purely diagonal.  For the SM quarks and 
charged leptons, their interactions are given by
\begin{eqnarray}
{\cal L}_{int} &=& g_D X_\mu \left[ -{3 \over 2} \bar{u}'_R \gamma^\mu u'_R 
+ {3 \over 2} \bar{d}'_R \gamma^\mu d'_R -{3 \over 4} \bar{c}'_R \gamma^\mu 
 c'_R + {3 \over 4} \bar{s}'_R \gamma^\mu s'_R \right. \nonumber \\  
&+& \left. 2 \bar{e}'_L \gamma^\mu e'_L + {5 \over 2} \bar{e}'_R 
\gamma^\mu e'_R + \bar{\mu}'_L \gamma^\mu\mu'_L + {5 \over 4} 
\bar{\mu}'_R \gamma^\mu \mu'_R \right],
\end{eqnarray}
where $u'_R,d'_R,c'_R,s'_R,e'_L,e'_R,\mu'_L,\mu'_R$ are not mass eigenstates, 
but linear combinations according to the diagonalizations of 
${\cal M}_u,{\cal M}_d,{\cal M}_l$ of Eqs.~(10) and (12).  This induces 
flavor-changing interactions such as $\mu \to eee$~\cite{n93}.  For $M_X$ 
of about 5 TeV, this means that the $\mu-e$ mixing should be less than about 
$10^{-2}$.  See Ref.~\cite{m20} for details.

Since there is no common Higgs vacuum expectation value which breaks both 
the SM gauge symmetry and $U(1)_X$, there is no $Z-X$ mixing at tree level, 
which is consistent with the phenomenological bound~\cite{pdg18} of about 
$10^{-4}$.

\noindent \underline{\it Concluding Remarks}~:~
The notion of hierarchical quark and lepton masses according to family is 
implemented by the sequential application of an unusual $U(1)_X$ gauge 
symmetry enabled by a right-handed lepton triplet instead of the usual 
right-handed neutrino singlet.  The Higgs sector consists of just the SM 
doublet and two $U(1)_X$ singlets with charges 1/4 and 3/4.  Neutrino 
masses come from Type III seesaw.  Higgs couplings to light quarks and 
leptons are predicted to differ from those of the SM.  Flavor-changing 
interactions from the $U(1)_X$ gauge boson are expected.

\noindent \underline{\it Acknowledgement}~:~
This work was supported in part by the U.~S.~Department of Energy Grant 
No. DE-SC0008541.

\baselineskip 20pt

\bibliographystyle{unsrt}

\end{document}